\documentclass[prl,twocolumn,groupedaddress,showpacs]{revtex4}

\usepackage{graphicx,amsmath}
\usepackage{feynmp}
\usepackage{color}

\begin{document}

\title{The Effects of Landau Level Mixing on the Fractional Quantum Hall Effect in Monolayer Graphene}

\author{Michael R. Peterson$^{1}$ and Chetan Nayak$^{2,3}$}
\affiliation{$^{1}$Department of Physics \& Astronomy, California State University Long Beach,  Long Beach, California 90840, USA}
\affiliation{$^{2}$Department of Physics, University of California, Santa Barbara, California 93106, USA}
\affiliation{$^{3}$Microsoft Research, Station Q, Elings Hall, University of California, Santa Barbara, California 93106, USA}

\begin{abstract}
We report  results of exact diagonalization studies of the spin- and valley-polarized fractional quantum Hall effect in the
$N=0$ and 1 Landau levels in graphene. We use an effective model that incorporates Landau level mixing to
lowest-order in the parameter $\kappa = \frac{e^2/\epsilon\ell}{\hbar v_F/\ell}=\frac{e^2}{\epsilon v_F\hbar}$
which is magnetic field independent and can only be varied through the choice of substrate.
We find Landau level mixing effects are negligible in the $N=0$ Landau level for $\kappa\lesssim 2$.
In fact,  the lowest Landau level projected Coulomb Hamiltonian is a better approximation to the real Hamiltonian for graphene than it is for semiconductor based
quantum wells. 
Consequently, the principal fractional quantum Hall states are expected in the $N=0$ Landau level over this
range of $\kappa$. In the $N=1$ Landau level,  fractional quantum Hall states are expected for a smaller range of $\kappa$
and Landau level mixing strongly breaks particle-hole symmetry producing qualitatively different results compared to the $N=0$ 
Landau level. At half-filling of the $N=1$ Landau level,
we predict the anti-Pfaffian state will occur for $\kappa \sim 0.25$-$0.75$. 
\end{abstract}

\date{\today}

\pacs{71.10.Pm, 71.10.Ca, 73.43.f}

\maketitle

\textit{Introduction}--The fractional quantum Hall effect (FQHE) occurs when electrons are confined to two-dimensions and placed in a uniform 
perpendicular magnetic field at electron densities $\rho$ such that the filling factor
$\nu=2\pi\ell^2\rho$ is a rational fraction ($\ell=\sqrt{\hbar c/eB}$  is the magnetic length) and the 
temperature is low (typically on the order of a Kelvin)~\cite{Tsui82}. For densities $\rho\sim 10^{11}$/cm$^2$,
the magnetic field strength $B$ must typically be from several Tesla up to tens of Teslas.
A plateau is observed in the Hall resistance with $R_{xy}=h/fe^2$, for a rational number $f$, along with
a concomitant vanishing of the longitudinal resistance $R_{xx}=0$.
Since the electrons in graphene move in a two-dimensional layer of negligible width and interact through
a Coulomb interaction with dielectric constant on the order of $1$, they would appear to realize a nearly perfect
setting for the FQHE.
Instead, the FQHE in graphene has remained puzzling~\cite{Dean11,Feldman12} since the initial experimental 
observations~\cite{Bolotin09,Du09}, even though early calculations predicted that the FQHE in 
graphene would be nearly identical to the FQHE in semiconductor heterostructures in the 
 lowest electronic Landau level (LL)~\cite{Apalkov06,Toke06,Nomura06,Goerbig06}.  
Despite the fact that the single particle dispersion is linear (relativistic) in graphene and quadratic in semiconductor heterostructures, 
the Haldane pseudopotentials in the $N=0$ LL for both systems are identical in the absence of LL mixing.

On closer inspection, LL mixing--not taken into account in previous theoretical studies--may be very different in graphene than in semiconductors.
Landau level mixing occurs when electrons in the fractionally filled $N^\mathrm{th}$ LL have a
substantial probability amplitude of making virtual transitions to higher and lower LLs.
This tendency is characterized by the ratio between the Coulomb interaction energy and 
the cyclotron energy, i.e., the LL mixing parameter $\kappa$ is defined as:
\begin{equation}
\kappa = \left\{%
\begin{array}{ll}
\frac{e^2/\epsilon\ell}{\hbar \omega}\sim\frac{2.5}{\sqrt{B[\mathrm{Tesla}]}} \mathrm{\;\;\;GaAs\; semiconductor}\\
 \frac{e^2/\epsilon\ell}{\hbar v_F/\ell}=\frac{e^2}{\epsilon v_F\hbar} \mathrm{\;\;\;\;\;\;\;\;\;graphene}
\end{array}%
\nonumber
\right.
\end{equation}
where $\omega=eB/mc$.  In semiconductors,  
$\kappa$ is inversely proportional to the magnetic field strength $B$ and therefore can, in principle, be made  small with a sufficiently large magnetic field. Traditionally, this was a primary motivation for ignoring LL mixing effects in these systems.
In graphene,  $\kappa$ has no magnetic field dependence and only depends 
on material properties, namely, the Fermi velocity $v_F$ and the dielectric constant $\epsilon$.  For a suspended graphene 
sheet $\kappa\approx 2.2$ and for graphene placed on  substrates such as SiO$_2$, $\kappa\approx 0.9$, 
or Boron Nitride, $\kappa\approx 0.5$-$0.8$~\cite{DasSarma-RMP2011,Peterson13b}.
Clearly, LL mixing cannot safely be ignored,
particularly in freestanding graphene where the FQHE was first experimentally observed.

Recently we constructed an effective Hamiltonian for the FQHE in graphene that fully incorporates 
Landau level mixing~\cite{Peterson13b}.  This effective Hamiltonian for 
electrons fractionally filling the $N^\mathrm{th}$ LL was produced by integrating out all
other LLs to first order in $\kappa$ (following Ref.~\onlinecite{Bishara09a}) and
is characterized by Haldane pseudopotentials
\begin{eqnarray}
H(\kappa)&=&\sum_{i<j}V_\mathrm{eff}(\kappa,|\mathbf{r}_i-\mathbf{r}_j|)+\sum_{i<j<k}V_\mathrm{3body}(\kappa,\mathbf{r}_i,
\mathbf{r}_j,\mathbf{r}_k)\nonumber\\
&=&\sum_{\alpha}V^{(2)}_\alpha(N,\kappa)\sum_{i<j}\hat{P}_m(m_{ij})\nonumber\\
 &&\hspace{1cm}+ \sum_\beta V^{(3)}_\beta(N,\kappa)\sum_{i<j<k}\hat{P}_{ijk}(m_{ijk})
\label{Heff}
\end{eqnarray}
where $\hat{P}_{ij}(m_{ij})$ and $\hat{P}_{ijk}(m_{ijk})$ project electrons $i$ and $j$ or $i, j$, and $k$ 
onto states with relative angular momentum $m_{ij}$ or $m_{ijk}$, respectively. 
$V^{(2)}_\alpha(N,\kappa)$ and $V^{(3)}_\beta(N,\kappa)$ are the $\kappa$ dependent
two and three-body effective Haldane pseudopotentials~\cite{Haldane83,Simon07c}.  (Similar to Ref.~[\onlinecite{Peterson13b}], we use planar geometry 
pseudopotentials throughout this work.) 
The expansion to lowest-order in $\kappa$ is especially interesting because, in addition to renormalizing
the two-body Coulomb interaction, it generates three-body terms that explicitly break particle-hole symmetry. 
The most important aspects of this effective Hamiltonian are: (i) in the $N=0$ LL the three-body terms
vanish due to particle-hole symmetry, which is an exact symmetry only in this LL, (ii) the two-body corrections are numerically 
small for $N=0$, and (iii) the size and character of the LL mixing corrections make the FQHE unlikely for $N\geq 2$.  
(See Ref.~[\onlinecite{Peterson13b}] for more details, especially Fig. 11.)  

In this work, we have performed numerical exact diagonalization of Eq.~(\ref{Heff}) in 
the spherical geometry in the $N=0$ and 1 LLs, focusing  
on filling factor $\nu=1/3$, $2/3$, and $1/2$. The first two are representative
of well-understood fractions in GaAs and are almost certainly Abelian~\cite{Laughlin83,Jain89,jain2007composite} while
the third is still not completely understood in GaAs but is suspected to be non-Abelian
in the $N=1$ LL~\cite{Moore91,Nayak96c,Nayak08}.
In our calculations $N_e$ electrons are placed on a spherical surface of radius $\sqrt{N_\Phi/2}$
with a radial magnetic field produced by a magnetic monopole of strength $N_\Phi/2$
at the center ($N_\Phi$ is required by Dirac to be an integer).  The relationship between the 
magnetic field strength and the number of particles is $N_\Phi = N_e/\nu - S$, where $S$ is a topological quantum number known as the ``shift"~\cite{Wen90b} 
and the filling fraction is $\nu=\lim_{N_e\rightarrow\infty} N_e/N_\Phi$.  A FQH  state will possess rotational invariance (total angular momentum $L=0$)
and an energy gap that remains finite in the thermodynamic limit.
Particle-hole symmetry plays a central role in the $N=1$ LL since the three-body terms that emerge from
LL mixing break this symmetry.  Hence, particle-hole conjugated states may have very different physics (the 
particle-hole conjugate relationship is found through $N_h= N_\Phi + 1 - N_e$.)

Since we are focusing on $\nu=1/3$, $2/3$, and $1/2$ we will compare the exact ground states of Eq.~(\ref{Heff}) with 
the Laughlin~\cite{Laughlin83} state at $\nu=1/3$ and its particle-hole 
conjugate at $\nu=2/3$ and the Moore-Read (MR) Pfaffian~\cite{Moore91} and anti-Pfaffian~\cite{Lee07,Levin07}
states at $\nu=1/2$.  These FQH states correspond to shifts of $S=3$ and $0$ for the Laughlin 1/3 and 2/3 states, respectively, 
and  $S=3$ and $-1$ for the MR Pfaffian and anti-Pfafffian, respectively.

We consider only fully spin- and valley-polarized states so our results apply to
experimental configurations in which spin- and valley-degeneracy are explicitly broken, for example, by the substrate or in bilayer graphene~\cite{Ki2014,Papic2014,Abergel09,Chakraborty13}.  However, the single particle dispersion in bilayer graphene is quadratic compared to the linear dispersion in monolayer graphene, thus, we caution the reader that our results might only be qualitatively applicable there.  Ref.~\onlinecite{Papic2014}  provided an explanation for the recent 
experimental observation of a 1/2-filled FQHE in bilayer graphene~\cite{Ki2014} but were unable to distinguish MR Pfaffian from anti-Pfaffian.  Perhaps our 
work can shed some light on that question.
Our results also apply to those states in which spin- and valley-polarization occurs spontaneously--or nearly spontaneously, since
weak SU(4) symmetry-breaking effects are present~\cite{Papic09,Amet2014}. It is beyond the scope of the
present work to study  physics that leads to spin- and valley-polarization or to study states not
fully polarized~\cite{Feldman13,Abanin13,Sodemann2014}.
The three-body terms in the Eq.~(\ref{Heff}), while not increasing the Hilbert space dimension, drastically 
decrease the sparsity of the Hamiltonian matrix.  While adding spin is possible and 
will be done in a future study~\cite{Kiryl}, adding spin \textit{and} valley degrees of freedom
is numerically prohibitive. It is likely only possible  to consider system sizes on the order of six particles
while including both effects. Hence, these two effects, in the 
context of exact diagonalization, will have to await further studies and/or further numerical and theoretical breakthroughs.

Before describing our  results, we describe the connection between the filing factor $\nu$ in our calculations and the 
observed Hall conductance $\sigma_{xy} = f(e^2/h)$.
We model electrons with filling fraction $\nu$ in the 
$N^\mathrm{th}$ LL by considering electrons at $\nu$ in the $N=0$ LL
and account for $N\neq 0$ by modifying the Haldane pseudopotentials appropriately.
Because of particle-hole symmetry about $N=0$, the Hall conductance $f$ is related to the Landau level $N$ and its fractional filling
$\nu$ according to $f = 4N - 2 + k+ \nu$. The factor of $4$ is due to the spin and valley degrees and
$k=0,1,2,3$ labels the possible spin and valley quantum numbers within a LL.
Therefore, our results for $\nu=1/3$ and $N=0$ apply to $f=-\frac{5}{3},-\frac{2}{3}, \frac{1}{3}, \frac{4}{3}$, and 
for $N=1$ the corresponding $f$s are $f=\frac{7}{3},\frac{10}{3},\frac{13}{3},\frac{16}{3}$. Meanwhile,
our results for $\nu=2/3$ and $N=0$ apply to $f=-\frac{4}{3}, -\frac{1}{3}, \frac{2}{3}, \frac{5}{3}$ and for 
$N=1$ the corresponding $f$s are $f=\frac{8}{3},\frac{11}{3},\frac{14}{3},\frac{17}{3}$.
Finally, $\nu=1/2$ corresponds, in the $N=0$ LL to $f=-\frac{3}{2}, -\frac{1}{2}, \frac{1}{2}, \frac{3}{2}$ and, in the $N=1$ LL,
to $f= \frac{5}{2}, \frac{7}{2}, \frac{9}{2}, \frac{11}{2}$.  On the hole side, for $N=-1$, one simply transforms all $f\rightarrow-f$.

\textit{Results}--We first report our results and then provide justification.  We conclude that (i) LL mixing has a large qualitative effect on the FQHE in the $N=1$ LL. 
The $\nu=1/3$ FQHE ($f=\frac{7}{3},\frac{10}{3},\frac{13}{3},\frac{16}{3}$)  survives even with strong LL mixing,
but the particle-hole conjugate state at $\nu=2/3$ ($f=\frac{8}{3},\frac{11}{3},\frac{14}{3},\frac{17}{3}$) does not.
We predict a FQHE state in the $\nu=1/2$ filled $N=1$ LL that is likely in the universality class of the anti-Pfaffian state~\cite{Lee07,Levin07}.    (ii) The FQHE in the $N=0$ LL of graphene is nearly identical to the FQHE in  the
$N=0$ LL in semiconductor heterostructures even in the presence of strong LL mixing.  Amusingly, the 
FQHE in the $N=0$ LL in graphene is \textit{more} like the minimal theoretical model than semiconductor systems:
graphene has no finite-thickness modification of the Coulomb potential, and LL mixing does not generate three-body terms as
a result of particle-hole symmetry.
As such, the $\nu=1/2$ case is found to be, as in semiconductor heterostructures, a 
composite fermion Fermi sea~\cite{Halperin93,Rezayi94,jain2007composite}.

\begin{figure}[t]
\begin{center}
\includegraphics[width=9cm,angle=0]{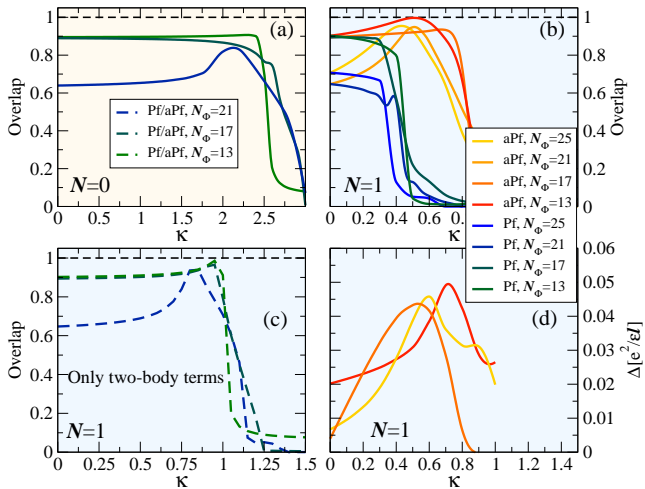}
\caption{(Color online) The  wave function overlaps between the exact ground state of Eq.~(\ref{Heff}) and the MR Pfaffian 
and anti-Pfaffian as a function of LL mixing ($\kappa$) 
for the (a) $N=0$ and (b) $N=1$ LLs.  The FQHE energy gap (exciton energy--far separated quasiparticle and quasihole) for the $N=1$ LL in units of $e^2/\epsilon\ell$ as a 
function of $\kappa$ are shown in (d).  Note that the $N_\Phi=21$ system is aliased with a 
composite fermion state at $\nu=4/9$, hence, these results are ambiguous and not included.  Finally, (c) shows wave function overlaps in the $N=1$ LL 
for the exact ground state of Eq.~(\ref{Heff}) excluding any 
particle-hole symmetry breaking three-body terms demonstrating its qualitative similarity with the $N=0$ LL.}
\label{fig-12}
\end{center}
\end{figure}

\textit{Graphene FQHE in half-filled Landau levels}--In Fig.~\ref{fig-12}(a)-(c) we show the numerical wave function overlap between 
the exact ground state of the effective Hamiltonian in Eq.~(\ref{Heff}) for $\nu=1/2$ in the $N=0$ and $N=1$ LLs and the 
Moore-Read Pfaffian ($N_\Phi=2N_e-3$) and anti-Pfaffian ($N_\Phi=2N_e+1$) wave functions as a function of the LL mixing 
parameter $\kappa$.  For $N=0$ we do not consider the overlap with the anti-Pfaffian since there 
are no particle-hole symmetry breaking three-body terms, i.e., the MR Pfaffian and anti-Pfaffian are degenerate.  For the lowest LL 
(Fig.~\ref{fig-12}(a)) the 
overlap is relatively insensitive to LL mixing until approximately $\kappa\sim2$ when it increases slightly before collapsing to zero.  In fact, this 
behavior, and others not shown, are consistent with previous results for $\nu=1/2$ in the lowest LL of semiconductor systems~\cite{Rezayi94}.  In 
contrast, in the $N=1$ LL (Fig.~\ref{fig-12}(b)), LL mixing increases the overlap between the ground state and the anti-Pfaffian to a 
maximum above 0.93 while the overlap with the MR Pfaffian monotonically decreases.  The latter phenomenon is the opposite of what happens in the case of GaAs in the N = 1 LL~\cite{Wojs10}.
The dramatic effect of the LL mixing induced three-body terms can 
be seen if one  considers only the two-body terms in Eq.~\ref{Heff}.  In that case, the behavior is qualitatively similar to the $N=0$ LL (Fig.~\ref{fig-12}(c)). 

Next we calculate the FQHE energy gap (for a presumed paired state) for a far-separated quasiparticle and quasihole pair (an exciton), which is the 
difference between the lowest energy at $L=N_e/2$ for $N_e/2$ even and $L=N_e/2-1$ for $N_e/2$ odd  and the absolute ground state at $L=0$. If the ground state does 
not have $L=0$ the gap is taken to be zero. This method
avoids some aliasing problems inherent in finite sized FQHE studies and is a useful
alternative to a computation comparing ground state energies for different values of flux $N_\Phi$~\cite{Morf02}.  But even with this method we still ignore the $N_\Phi=21$ state 
when calculating the gap since it is aliased with an abelian composite fermion state~\cite{Jain89,jain2007composite}.
Interestingly, the FQHE energy gap is a non-monotonic function of $\kappa$; a maximum is obtained around $\kappa\sim 0.5$-$0.7$ (Fig.~\ref{fig-12}(d)).

\begin{figure}[]
\begin{center}
\includegraphics[width=9cm,angle=0]{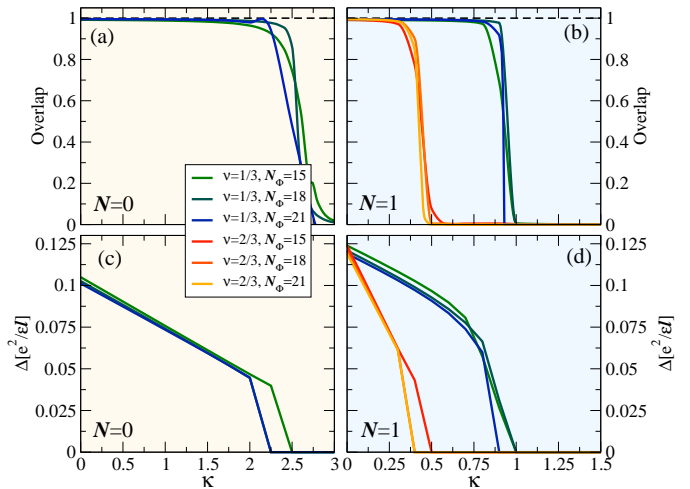}
\caption{(Color online) The  wave function overlaps between the exact ground state of Eq.~(\ref{Heff}) and the Laughlin state as a 
function of LL mixing ($\kappa$) 
for the (a) $N=0$ and (b) $N=1$ LLs.  Panels (c) and (d) show FQHE excitation gaps (defined in the text) for the $N=0$ and $N=1$ LL in units of $e^2/\epsilon\ell$ as a 
function of $\kappa$.}
\label{fig-13}
\end{center}
\end{figure}

\textit{Graphene FQHE in $1/3$ and $2/3$ filled Landau levels}--Figures~\ref{fig-13}(a) and (b) show the overlap between the Laughlin wavefunction or its particle-hole conjugate and the exact ground state of Eq.~(\ref{Heff}) at $\nu=1/3$ or $2/3$ in the $N=0$ and $N=1$ 
LLs, i.e., $N_\Phi=3N_e-3$ or $N_\Phi=3N_e/2$, respectively.  Again, for $N=0$ we only show overlaps with the Laughlin wave function 
at $1/3$ since there are no three-body terms present to break particle-hole symmetry.  The overlap remains very large ($\sim 0.99$) 
until $\kappa\approx 2$ where it abruptly drops to zero.  In the $N=1$ LL, we find that LL mixing breaks particle-hole symmetry for modest 
values of $\kappa$ and the overlaps with the Laughlin wavefunction at $1/3$ and $2/3$ markedly diverge; the $1/3$ overlap remains $\sim 0.99$ until $\kappa\sim 1$ 
while the $2/3$ overlap remains large only until $\kappa\sim0.4$.  

In Fig.~\ref{fig-13}(c) and (d) we calculate the FQHE energy gap as the difference between the lowest energies at $L=N_e$ and $L=0$ (unpaired excitation).    
The gap decreases monotonically with $\kappa$ for $N=0$ until collapsing to zero around $\kappa\sim2$ coinciding with the $\kappa$ where the overlap vanishes.  For $N=1$ we find
 that the gap decreases monotonically with $\kappa$ until the overlaps and gaps  collapse to zero simultaneously.  The 1/3 state survives much 
 stronger LL mixing (to $\kappa\sim 1$) while the 2/3 state does not (the overlap and gap vanish at $\kappa\sim 0.4$).

\begin{figure}[]
\begin{center}
\includegraphics[width=8cm,angle=0]{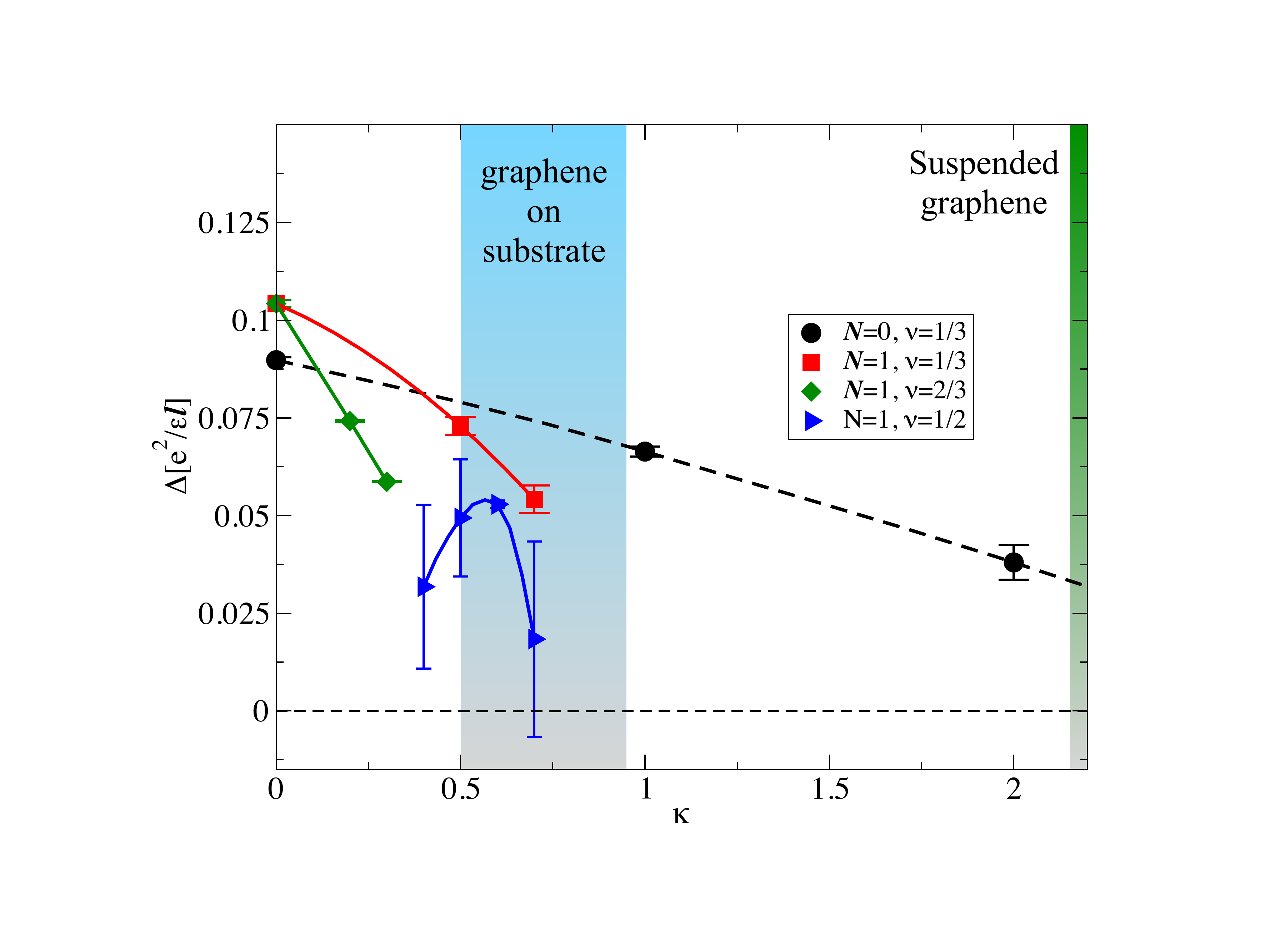}
\caption{(Color online) FQHE energy gaps extrapolated to the thermodynamic limit as functions of $\kappa$ for $\nu=1/3$, and $2/3$ in the $N=0$ and $N=1$ LLs and $\nu=1/2$ in the $N=1$ LL.  The 
1/3 and 2/3 gaps decrease with $\kappa$ while at $\nu=1/2$ the gap shows a maximum in $\kappa$ at approximately the same value where the overlap with the anti-Pfaffian is maximum (see Fig.~\ref{fig-12}(b)).  The lines are merely a guide to the eye and the colored boxes indicate the approximate range of $\kappa$ for graphene on a substrate (blue) and suspended graphene (green).}
\label{fig-13-gaps}
\end{center}
\end{figure}

\textit{FQHE gaps in the thermodynamic limit and experimental comparison}--Last, we show the FQHE gaps extrapolated to the thermodynamic limit for $\nu=1/3$, $2/3$, and $1/2$ in the $N=0$ and $N=1$ LL (Fig.~\ref{fig-13-gaps}). 
If $\kappa=e^2/\epsilon v_F\hbar$ is varied in an experiment, by changing the
dielectric constant $\epsilon$, then the energy gap must be plotted in units of the changing scale $e^2/\epsilon\ell$ (or else the scale must be held constant by simultaneously
varying $B$).
Consider two examples:  (1) In Ref.~\onlinecite{Bolotin09},  the energy gap 
of $f=1/3$ was measured to be $\Delta_\mathrm{exp}\sim60\mathrm{K}$ at $B=14\mathrm{T}$ in a suspended sample 
with $\kappa=2.2$.  As shown in Fig.~\ref{fig-13-gaps},
the calculated gap is $\Delta_\text{calc}\sim 0.035 e^2/\epsilon\ell$.  Since $\epsilon=1$ for a suspended sample, this
corresponds to $\Delta_\text{calc}\sim 85\mathrm{K}$, which is differs from the experimental result by a factor of approximately 1.5.  Considering that we have neglected the
effects of disorder, this is an encouraging result.  (2) In Ref.~\onlinecite{Dean11}, $\Delta_\mathrm{exp}\sim12\mathrm{K}$ at $B\sim28\mathrm{T}$ for $f=4/3$.  
Taking $\kappa=0.5$ and $\epsilon\sim 5$, our calculations yield $\Delta\sim 50\mathrm{K}$ -- a factor of about 5 too large.  
Perhaps this poorer estimate  stems from different disorder characteristics of graphene on a substrate and/or the neglect of spin and valley degrees of freedom~\cite{Abanin13}.

\textit{Conclusions}--(i) When spin- and valley-degeneracy are broken, the 
FQHE in the $N=0$ LL of graphene is expected to be nearly identical to the $B\rightarrow\infty$ minimal model of the FQHE (pure Coulomb Hamiltonian) as long 
as $\kappa\leq2$.  Thus, all of the known results  in the $N=0$ LL for semiconductor systems transfer to graphene nearly perfectly 
even in the presence of LL mixing. (ii) The FQHE is expected in the $N=1$ LL for moderate values of $\kappa$ -- which might be expected on Boron Nitride and SiO$_2$ substrates 
but not in suspended samples where LL mixing is too strong.  We find strong particle-hole symmetry breaking in the $N=1$ LL,
leading to  stark differences between the 1/3 FQHE and the particle-hole symmetric partner at 2/3, i.e., the 1/3 state would exist in a system with $\kappa=0.7$ and the 
2/3 might not.  (iii) Intriguingly, we find the anti-Pfaffian state to be stabilized in the $N=1$ LL for moderate values of $\kappa\sim 0.25-0.75$.  The 
MR Pfaffian, on the other hand, is disfavored by LL mixing.

While our results predict that the $\nu=1/3$ and $2/3$ states will be related by symmetry in the $N=0$ LL and the latter will be suppressed
in the $N=1$ LL, the experimental situation is  more complicated. In the $N=0$ LL, odd-numerator states are generally suppressed. However,
this is likely due to the presence of low-energy Skyrmion excitations in a spontaneously spin- and valley-polarized state~\cite{Sondhi1993,MacDonald1996,Abanin13}; 
however, LL mixing is known to generally effect Skyrmion excitations~\cite{Melik99,Mihalek00}.
When the valley symmetry is explicitly broken, for example, by a substrate or with an applied
electric field in a bilayer system, odd numerator states are strengthened. 
In the $N=1$ LL, the $\nu=7/3$ state is stronger than the $\nu=8/3$ state, in agreement with our calculations, yet the $\nu=11/3$ and 10/3 states have comparable gaps~\cite{Dean11}, similar to the experimental observations of the $\nu=7/3$ and 8/3 gaps in GaAs semiconductors.  Last, we note that a 1/2-filled FQH state has only been experimentally observed in bilayer graphene~\cite{Ki2014} and so far no experimental groups have definitively observed a 1/2-filled state in the N=1 
in monolayer graphene unlike in GaAs.  We hope our work will motivate more experimental investigations in graphene.

The FQHE in graphene provides a diverse playground where interplay between LL mixing, disorder, spin and valley degrees of freedom lead to rich
and surprising physics.  In this work, we have focused on the  effects of LL mixing.

\textit{Acknowledgments}--We are grateful to DARPA QuEST and Microsoft for
funding and MRP thanks California State University Long Beach Start-up Funds and the Office of Research and Sponsored Programs at CSULB.   We acknowledge helpful discussions 
with Francois Amet, Cory Dean, Jainendra Jain, Philip Kim, Kiryl Pakrouski, and Andrea Young.

\end{document}